\documentclass[journal = nalefd, manuscript = letter, layout = twocolumn]{achemso}
\usepackage{amsmath}
\pdfoutput=1
\let\oldmaketitle\maketitle
\let\maketitle\relax

\author{Francesco Basso Basset}
\email{f.bassobasset@campus.unimib.it}
\affiliation
{L-NESS and Dipartimento di Scienza dei Materiali, Universit\`a degli Studi di Milano-Bicocca, Via Cozzi 55, I-20125 Milano, Italy}
\alsoaffiliation
{Institute of Semiconductor and Solid State Physics, Johannes Kepler University, Altenbergerstra{\ss}e 69, Linz 4040, Austria}
\author{Sergio Bietti}
\affiliation
{L-NESS and Dipartimento di Scienza dei Materiali, Universit\`a degli Studi di Milano-Bicocca, Via Cozzi 55, I-20125 Milano, Italy}
\author{Marcus Reindl}
\affiliation
{Institute of Semiconductor and Solid State Physics, Johannes Kepler University, Altenbergerstra{\ss}e 69, Linz 4040, Austria}
\author{Luca Esposito}
\affiliation
{L-NESS and Dipartimento di Scienza dei Materiali, Universit\`a degli Studi di Milano-Bicocca, Via Cozzi 55, I-20125 Milano, Italy}
\author{Alexey Fedorov}
\affiliation
{L-NESS and CNR-IFN, via Anzani 42, I-22100 Como, Italy}
\author{Daniel Huber}
\author{Armando Rastelli}
\affiliation
{Institute of Semiconductor and Solid State Physics, Johannes Kepler University, Altenbergerstra{\ss}e 69, Linz 4040, Austria}
\author{Emiliano Bonera}
\affiliation
{L-NESS and Dipartimento di Scienza dei Materiali, Universit\`a degli Studi di Milano-Bicocca, Via Cozzi 55, I-20125 Milano, Italy}
\author{Rinaldo Trotta}
\email{rinaldo.trotta@jku.at}
\affiliation
{Institute of Semiconductor and Solid State Physics, Johannes Kepler University, Altenbergerstra{\ss}e 69, Linz 4040, Austria}
\author{Stefano Sanguinetti}
\affiliation
{L-NESS and Dipartimento di Scienza dei Materiali, Universit\`a degli Studi di Milano-Bicocca, Via Cozzi 55, I-20125 Milano, Italy}
\alsoaffiliation
{L-NESS and CNR-IFN, via Anzani 42, I-22100 Como, Italy}

\title{High-yield fabrication of entangled photon emitters for hybrid quantum networking using high-temperature droplet epitaxy}

\begin{document}

\twocolumn[
\begin{@twocolumnfalse}
	\oldmaketitle
		\begin{abstract}
			Several semiconductor quantum dot techniques have been investigated for the generation of entangled photon pairs. Among the other techniques, droplet epitaxy enables the control of the shape, size, density, and emission wavelength of the quantum emitters. However, the fraction of the entanglement-ready quantum dots that can be fabricated with this method is still limited to around 5\%, and matching the energy of the entangled photons to atomic transitions---a promising route towards quantum networking---remains an outstanding challenge.
			
			Here, we overcome these obstacles by introducing a modified approach to droplet epitaxy on a high symmetry (111)A substrate, where the fundamental crystallization step is performed at a significantly higher temperature as compared to previous reports. Our method drastically improves the yield of entanglement-ready photon sources near the emission wavelength of interest, which can be as high as 95\% due to the low values of fine structure splitting and radiative lifetime, together with the reduced exciton dephasing offered by the choice of GaAs/AlGaAs materials. The quantum dots are designed to emit in the operating spectral region of Rb-based slow-light media, providing a viable technology for quantum repeater stations.
			
				
		\end{abstract}
\end{@twocolumnfalse}
]

\newpage

As a part of the ongoing effort to develop practical quantum technologies, the search for a suitable entangled photon source is an active research direction because such sources play an important role in key quantum communication protocols and some quantum computation approaches\cite{Kimble-2008,Pan-2012}. Most importantly, the development of such sources is a fundamental requirement for the realization of repeaters capable of transferring quantum entanglement over long distances.

Epitaxial quantum dots (QDs) are a promising alternative to parametric down-converters due to their ability to generate photons on-demand with high efficiency and their compatibility with semiconductor foundries\cite{Benson-2000,Lu-2014}. To use QD entanglement resources in practical technologies, two main roadblocks have to be overcome. The first is related to the difficulty of consistently finding emitters that can generate highly entangled photon pairs. The second concerns the wavelength of operation of the quantum source, which must be compatible with other components of a quantum network, such as storage media and detectors. In this work, we search for the best approach to face these challenges.

To achieve reproducible entangled photon generation, it is necessary to deal with the in-plane anisotropies in the confinement potential that induce a fine structure splitting (FSS) between the bright exciton states through the electron-hole exchange interaction\cite{Bester-2003,Seguin-2005}. Moreover, methods to alleviate exciton dephasing caused by the fluctuating magnetic fields produced by the QD nuclei must be taken into account.\cite{Hudson-2007}

Dephasing by nuclear magnetic fields depends mainly on the choice of the material. Among the various systems proposed to date\cite{Treu-2012,Mano-2010,Huo-2013,Seguin-2005,Yerino-2014,Juska-2013,Versteegh-2014}, GaAs QDs stand out as the best option. In contrast to the standard In(Ga)As QDs obtained by the Stranski-Krastanow method, GaAs QDs are weakly affected by exciton spin scattering due to their low nuclear magnetic moments. Indeed, a recent report\cite{Huber-2017} has shown unprecedented high levels of entanglement and indistinguishability in photon pairs generated from GaAs/AlGaAs nanostructures fabricated by droplet etching.

Here we want to draw attention to a different growth strategy based on droplet epitaxy (DE). The DE method for the fabrication of QDs is based on the sequential deposition of group III (Ga, In, Al) atoms at controlled temperature and flux to form nano-droplets on the surface, and of group V (As, P, Sb, N), to crystallize the droplets into nano-islands\cite{Sanguinetti-2013}. This technique presents some appealing advantages as compared to droplet etching, such as the much wider control over the spatial density of emitters\cite{Heyn-2007} and their shape\cite{Bietti-2015}. Moreover, this growth scheme is compatible with different materials so that it can be employed to fabricate emitters for a broad spectral range, notably also in the conventional telecom band\cite{Liu-2014,Skiba-Szymanska-2017}.

GaAs QDs grown by DE have already been proved to yield polarization-entangled photons with a very high fidelity\cite{Kuroda-2013}, without either the temporal post-selection or the external tuning\cite{Stevenson-2006,Bennett-2010,Trotta-2016,Muller-2009}. This was enabled by collectively improving the in-plane symmetry of the as-grown QDs by fabrication on a (111)-oriented substrate\cite{Singh-2009,Schliwa-2009}.
Despite the high potential of this fabrication method, DE is still quite far from meeting the fundamental requirements for the practical realization of a hybrid semiconductor-atomic quantum network.
First, the average FSS value is still too high and gives rise to a minor fraction of only approximately 5\% of entanglement-ready emitters, so that finding an emitter with good performance requires large-area and time-consuming scans.
Additionally, it is desirable for the emission wavelength to be matched with an atomic-based optical slow medium, Rb being the natural choice for the GaAs/AlGaAs system\cite{Akopian-2011}. Through the control of the shape and barrier composition, we aim to reduce the confinement potential and achieve a practical density of single photon emitters around 780 nm. When this requirement is satisfied, a weak external field can be used to tune the exciton emission within the hyperfine splitting of the $^{87}$Rb D$_2$ transitions to achieve the delay or storage of a polarization qubit\cite{Akopian-2011,Trotta-2016,Huang-2017,Patel-2010}.
Finally, a long-standing drawback of this growth technique is the low substrate temperature during the formation of the nanostructures and the surrounding barrier. This places a limit on the crystalline and optical quality of the material, which can only be partially overcome with an annealing process\cite{Mantovani-2004,Mano-2009}. 

Previous research in this direction has only taken its first steps on (111)-grown samples\cite{Jo-2012} and here we address all these issues by introducing a DE-QDs growth regime where the substrate temperature during the Ga droplet crystallization by As supply is increased by more than 300$^{\circ}$C with respect to previous reports\cite{Bietti-2015,Mano-2010}.

Extensive high-resolution single-dot spectroscopy experiments revealed reduced spectral wandering and successful control over the emission wavelength. We finally obtained a higher than 95\% fraction of the QDs emitting in the strategic wavelength range that shows compliance with the criteria for polarization-entangled photon generation. This is mainly due to the very low average value of FSS and fast radiative recombination. The demonstration of polarization-entangled photon emission is achieved via a two-photon resonant excitation scheme, which is employed here for the first time for DE nanostructures.

Samples of Ga droplets and GaAs QDs were fabricated on a GaAs (111)A substrate and embedded in an AlGaAs barrier. The detailed growth procedure and sample structure are reported in the Samples fabrication section.
First, metallic Ga nanodroplets are deposited in the absence of As supply and then are crystallized into GaAs with the rapid exposure to an intense As flux. In the conventional DE process, the crystallization of the Ga droplets into tridimensional QDs is achieved by controlling Ga diffusion length through substrate temperature and As pressure (low T and high As flux) so as to reduce the probability of Ga atoms migrating out of the droplet and binding to the As adsorbed by the surrounding AlGaAs surface. Therefore, the low temperature of approximately 200$^{\circ}$C required for the crystallization step has been the distinctive feature of this procedure. As the temperature is increased, the thermodynamically favorable planar growth of GaAs on AlGaAs becomes more dominant. In a recent work\cite{Bietti-2015}, we investigated this process in detail on standard (100) substrates and showed that island formation is observed up to 250$^{\circ}$C.

In this work, we present a different quantum dot fabrication regime by DE in which Ga droplets are crystallized, and the subsequent barrier layer is deposited at a high substrate temperature, close to the 520$^{\circ}$C used for the growth of high quality GaAs and AlGaAs on (111)A. This is expected to improve the crystallinity of the QDs by reducing the concentration of point defects typical of low-temperature growth\cite{Watanabe-2001,Mantovani-2004}. These defects are likely to cause spectral wandering and act as non-radiative recombination channels\cite{Mano-2009}.

This regime becomes accessible due to the specific choice of the (111)A substrate orientation. The droplet epitaxy growth proceeds via the balance of two competitive mechanisms activated by the As flux within and in the neighborhood of the droplet \cite{Bietti-2013,Reyes-2013,Somaschini-2009}: 1) a three-dimensional growth mode due to the direct incorporation of the As into the droplet itself at vapor-liquid interface and the subsequent crystallization at the droplet liquid-solid contact surface (vapor-liquid-solid growth mode); 2) a two-dimensional growth in the surrounding areas of the droplet due to metal diffusion from the droplet perimeter and its reaction with the impinging As flux. The balance between the two growth modes is set by the diffusivity of the metal species and by the flux and sticking coefficient of As. The dominance of the first leads to the formation of QDs \cite{Reyes-2013}. We have observed that the (111)A surface permits the crystallization under the As supply of the Ga nano-droplets into QDs even at a relatively high temperature (around 500$^{\circ}$C). Such behavior may originate from the shorter surface lifetime of As$_4$ on these surface orientations compared to the usual (001) surface\cite{Sato-1994}. The low As residence time limits the reactivity of the surface toward the bulk incorporation of the Ga adatoms that detach from the droplet perimeter during the crystallization and allows for a large fraction of the Ga stored in the droplet to crystallize in place.

\begin{figure}[ht]
	\includegraphics[width=8.4cm]{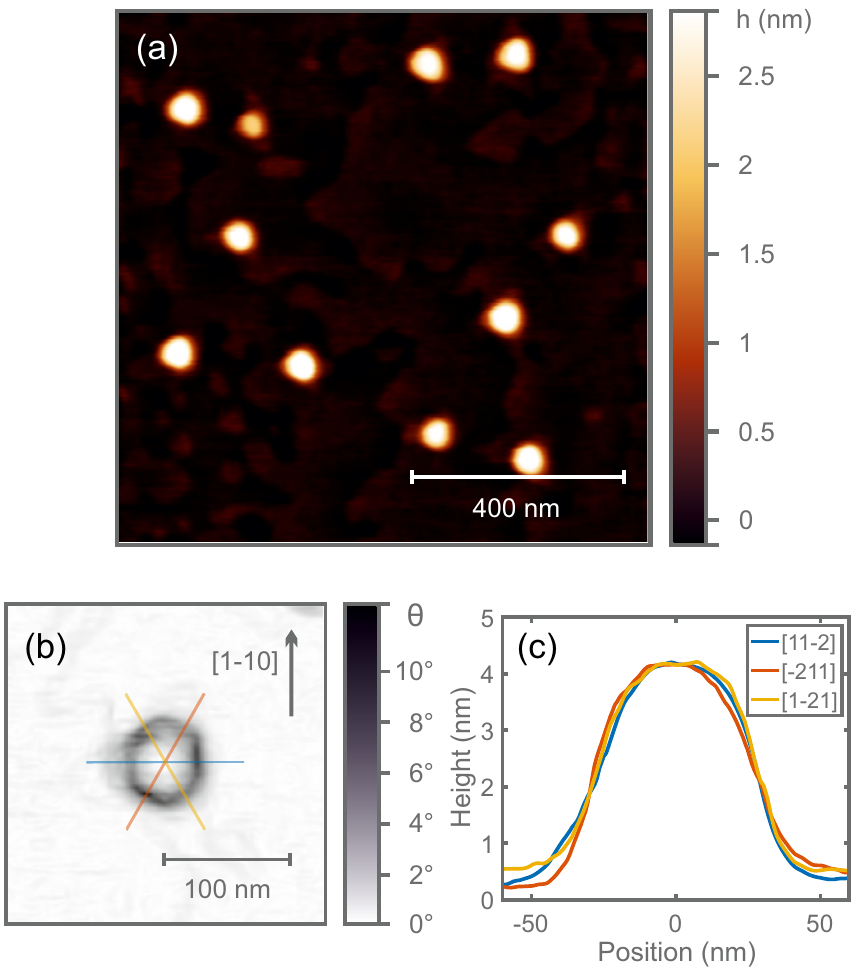}
	\caption{\label{fig:AFM} AFM scan of a QD sample. a) 1 $\mu$m $\times$ 1 $\mu$m map. b) Close-up of a typical single QD, map of the inclination angle $\theta $ with respect to the (111)A plane. c) Height profiles taken along [11-2] and equivalent crystallographic directions, following the colored lines in panel b).}
\end{figure}

The growth parameters were optimized to obtain a low density of emitters suitable for single dot studies together with size tailoring to achieve emission around the target wavelength of Rb atomic resonances.
The droplet sample was characterized by atomic force microscopy (AFM).  Control over droplet formation leads to the density of the Ga droplets of $9.1\times 10^{8}$ cm$^{-2}$. The droplets show the shape of spherical caps with the mean base diameter of 50.4 nm and mean height of 7.4 nm. The volume of the deposited Ga matches the volume of the droplets.

The formation of the nanocrystals after the crystallization step was evaluated by AFM on uncapped samples. The QD density and average QD volume are in good agreement with the droplet values (corrected by the factor taking into account the transformation of liquid Ga into GaAs in the case of the volume). While keeping a high substrate temperature, the As flux during crystallization was adjusted in order to obtain the desired geometry of the QDs. A quite large size, with the mean base diameter and height of 70 and 4 nm, respectively, and a low Al content in the barrier layer were employed to shift excitonic lines at longer wavelengths than those obtained in previous attempts\cite{Mano-2010}.
Figure~\ref{fig:AFM} reports the typical morphology of the nanostructures immediately after the crystallization. The base is hexagonal as shown in panel (a). In particular, a hexagonal truncated pyramid shape is observed in the more detailed map of QD-plane inclination shown in panel (b). A different color scale (shown in Supporting Information) would also reveal the presence of a very thin broader triangular base, which is likely due to the As incorporation outside the droplet and close to its perimeter. In the regions where the Ga adatom concentration is sufficiently high\cite{Bietti-2014}, layer-by-layer growth is promoted despite the low residence time of the As atoms. Fig.~\ref{fig:AFM}(c) shows mid-section height profiles collected across three directions which are equivalent according to $C_{3v}$ symmetry. The comparison highlights the high degree of the in-plane symmetry. We expect that the QD morphology will be maintained after capping as well, showing sharp interfaces \cite{Keizer-2010,Ha-2015}. This is due to the low concentration of point defects at the relatively high crystallization temperature of the QDs and the low interdiffusivity of Al and Ga species at the interface between the GaAs QD and the AlGaAs barrier \cite{Schlesinger-1986,Bracht-1999}.

\begin{figure}[ht]
	\includegraphics[width=8.4cm]{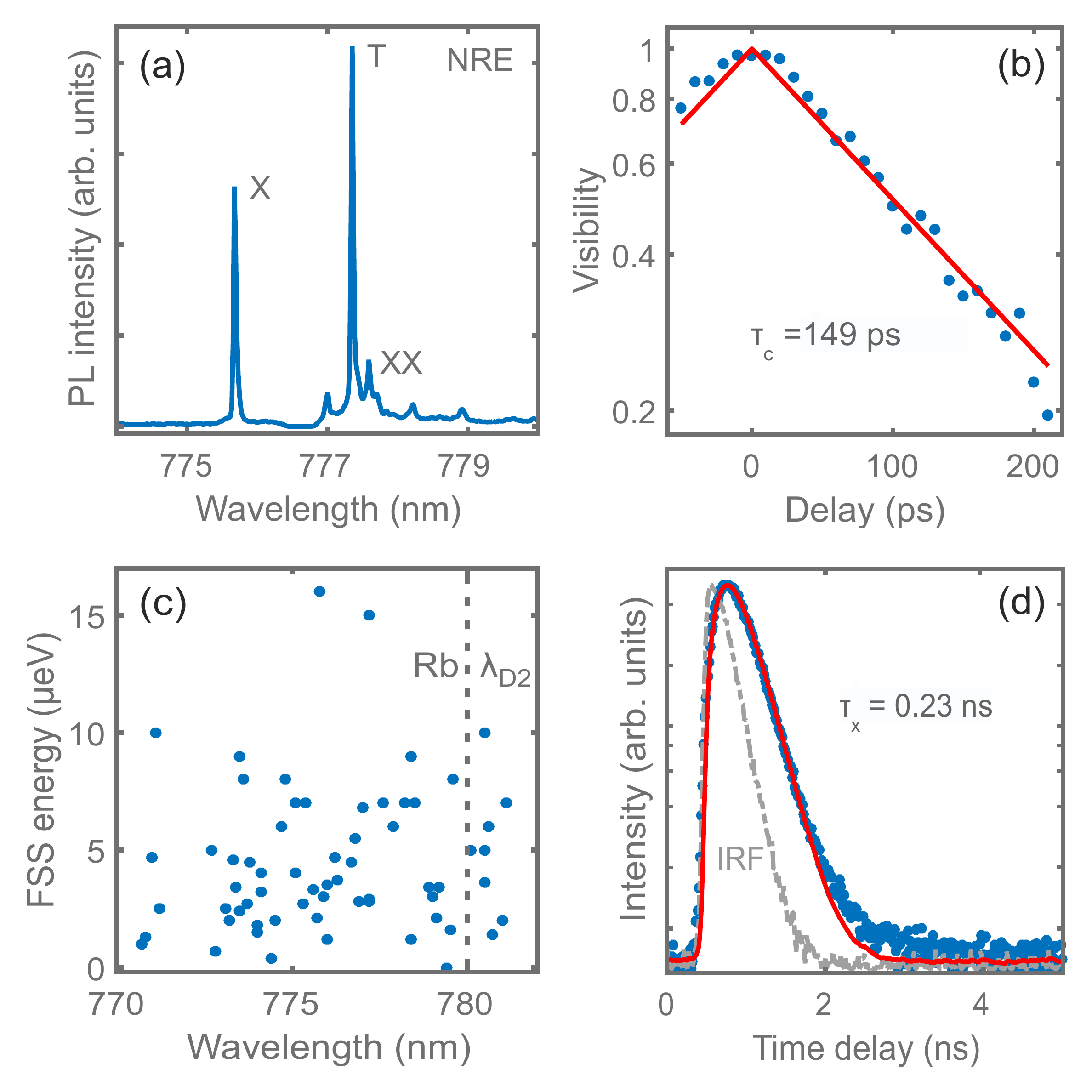}
	\caption{\label{fig:microPL_eV} a) Emission spectrum of a single QD under above-barrier excitation.  b) Interference visibility of a neutral exciton line from a selected QD (blue dots) fitted with exponential decay (red continuous line). c) FSS values measured on several QDs emitting at different wavelengths. d) Time decay of the exciton PL emission (blue dots) fitted (red continuous line) with exponential decay convoluted with the instrument response function (IRF, gray dashed line)}
\end{figure}

A direct assessment of the optical quality and the investigation of the electronic structure of these nanostructures required an in-depth analysis through single dot PL.

Figure~\ref{fig:microPL_eV}(a) shows a typical emission spectrum of our QDs under above-barrier excitation. It consists of an intense and isolated line attributed to the recombination of the neutral excitons (X) and is accompanied by several partially overlapping emissions at higher wavelengths. These lines are due to other excitonic complexes with positive binding energies, namely the singly charged excitons (T) and the biexciton (XX), together with other charged multiexcitonic states. The main peaks in the spectrum are labeled according to polarization-resolved measurements. This attribution is also supported by a power-dependence analysis (see Supporting Information) and it is consistent with previous experimental studies \cite{Abbarchi-2008} as well as with atomistic many-body pseudopotential calculations\cite{Luo-2011} for GaAs/AlGaAs DE QDs.

Contributions from several multiexcitonic complexes persist even at moderate and low excitation powers, possibly because a deep potential barrier in a weak lateral confinement regime leads to a large number of confined levels and thus a high probability of finding a carrier in an excited state or an extra charge captured by the tunneling from the energetically aligned defect states. An effective negative QD charging may strongly reduce the number of useful biexciton-exciton recombination events, a suggested by recent observations on pyramidal In$_{0.25}$Ga$_{0.75}$As QDs \cite{Juska-2015}. The presence of several spectrally overlapping multiexcitonic peaks has already been reported for the GaAs/AlGaAs nanostructures emitting at similar wavelengths but grown with a different epitaxial technique\cite{Jahn-2015}. However, it has also been shown that the biexciton radiative recombination can be selectively pumped by resorting to resonant excitation schemes\cite{Huber-2017}. This solution can be successfully applied to our DE QDs, as we will discuss below.

The linewidth of the neutral exciton line at low temperature is usually broadened by spectral diffusion\cite{Uskov-2001}. While it is hard to complete suppress this contribution in practice, we observed narrow emission lines on the samples with 15\% Al content in the barrier and 500$^{\circ}$C temperature for the droplet crystallization. A top barrier thickness above 100 nm is crucial for suppressing spectral diffusion from fluctuating surface charges\cite{Ha-2015}. We report that the majority of the neutral exciton lines have a linewidth below the resolution of the experimental setup (40 $\mu$eV).

To overcome the limited spectral resolution, we performed a series of coherence time measurements using a Michelson interferometer. Several neutral exciton lines were investigated and an example of the visibility decay as a function of the time delay is shown in Fig.~\ref{fig:microPL_eV}(b). The data are well-reproduced by a model which assumes a Lorentzian line broadening, even if a Gaussian contribution due to spectral wandering is always present, as commonly reported under an above-barrier excitation\cite{Berthelot-2006, Kuhlmann-2013}.
The Lorentzian fit immediately yields the exciton coherence time which can be easily translated into spectral broadening\cite{Kammerer-2002}. We report the average exciton zero-phonon linewidth of 15 $\mu$eV and the best value of 9 $\mu$eV, which is an improvement over the state of the art for DE QDs\cite{Mano-2009,Abbarchi-2011,Jo-2012}, namely the average of 35 $\mu$eV and the best value of 16 $\mu$eV. This result is most likely ascribed to the higher substrate temperature during the crystallization of the droplet and deposition of the barrier layer surrounding the QD, which therefore provides better material quality than the approaches relying on post-growth annealing\cite{Mantovani-2004,Mano-2009}.

While these values still do not reach the Fourier limit, which we will show to be 3 $\mu$eV at most, we recall that the reported measurements are performed under above barrier excitation and resonant excitation schemes might be required to suppress the charge noise\cite{Bayer-2002,Kuhlmann-2013}. However, it is important to point out that such a small level of dephasing is not expected to affect the entanglement fidelity\cite{Hudson-2007}.

As required for the construction of a Rb-based artificial-natural atomic interface, a fraction of the emitters matches the 780-nm spectral window (see Supporting Information for ensemble PL data). Spatial mapping over a 100 x 100 $\mu m^2$ area revealed more than 50 emitters with the neutral exciton line with a spectral distance of less than 2 nm from 780 nm. In this wavelength range, a simple external tuning technique would certainly allow precise matching with the Rb transitions. If required, the spatial density of the nanostructures can be further modified during the Ga droplet deposition step independently of their geometry\cite{Heyn-2007,Ohtake-2015}. Both the ensemble and individual QD optical properties have been probed across the sample on the length scale of the order of 10 mm and showed good uniformity.

An estimate of the FSS is obtained by mapping the energy position of the exciton and biexciton lines at different linear polarization angles\cite{Bayer-2002} (see Supporting Information for a typical measurement). The measured FSS values for the exciton lines above 770 nm are reported in Fig.~\ref{fig:microPL_eV}(c). The average FSS is strongly reduced compared to the DE on (100) substrates\cite{Abbarchi-2008} with the very low average value of 4.5 $\mu$eV and the standard deviation of 3.1 $\mu$eV. Moreover, the emitting dipole shows no preferential in-plane orientation, which is the direct consequence of the improved in-plane symmetry of the QDs presented here (see Fig.~\ref{fig:AFM}). Unlike for the (100) orientation, where an anisotropy in Ga diffusivity systematically causes an in-plane elongation, the nanocrystals grown on a (111)A substrate show the $C_{3v}$ symmetry required to achieve vanishing FSS\cite{Singh-2009,Schliwa-2009}. The average FSS value is also more than halved compared to the best values achieved to date using DE on (111)A substrates\cite{Kuroda-2013}. A possible explanation for this is that the symmetry of the confinement potential is higher in our QDs, because they are thicker and their overall shape is less affected by the accidental presence of the underlying monolayer step fluctuations of the substrate. In addition, it has been shown experimentally and theoretically that for a given QD shape, the FSS decreases with increasing dot size and hence decreasing confinement\cite{Huo-2014}. This is because the strength of the exchange interaction responsible for the FSS decreases as the carrier wavefunctions become more delocalized in large QDs (in our case, also due to the reduced band-offset between barrier and QD material). More generally, this result represents the state of the art for epitaxial systems on which entangled photon emission has been observed without the need for external tuning \cite{Kuroda-2013,Huo-2013,Juska-2013,Versteegh-2014,Young-2005,Keil-2017}.

In addition to the FSS values, the lifetime of optical transitions is another important parameter for the degree of entanglement. More specifically, the fidelity to the expected Bell state as measured in a time-average experiment depends dramatically on the ratio between the FSS and the exciton lifetime\cite{Huber-2017}. Therefore, we performed time-resolved measurements on the same sample. Figure~\ref{fig:microPL_eV}(d) shows the time decay of the PL intensity of a neutral exciton line under the above-barrier excitation. The excitation power was tuned to below the saturation level of the exciton in order to prevent the band filling effects\cite{Raymond-1996}. In such conditions, the experimental data can be described by the convolution of the instrument response function with a single exponential decay. We approximate the radiative lifetime to the total decay time of the system because in a high quality epitaxial QD at low temperature, non-radiative mechanisms are expected to be negligible\cite{Tighineanu-2013,Jahn-2015}. We expect an overestimation caused by the contribution to the measured total decay time from the relaxation processes that populate the exciton, so that faster radiative recombination can be achieved by resonant excitation (see below).
The radiative lifetime was evaluated for a series of several different QDs and the low average value of 300 ps (with the standard deviation of 60 ps, and 20 ps uncertainty for a single measurement) was obtained. This quantity depends rather weakly on the emission wavelength (as shown in the Supporting Information). Our values are shorter than those typical of the In(Ga)As QDs \cite{Dalgarno-2008,Langbein-2004,Trotta-2015-NL,Juska-2013}, dots embedded in InP nanowires\cite{Versteegh-2014} and previous reports of DE GaAs QDs\cite{Mano-2009,Kuroda-2013}, and are close to the best figures measured on GaAs QDs fabricated by droplet etching under quasi-resonant excitation\cite{Akopian-2013}.  

Due to the short radiative lifetime and low FSS, the conditions for polarization-entangled photon emission can be readily met. As mentioned above, once the values of the lifetime and FSS for each specific QD are known, it is possible to evaluate the expected entanglement fidelity quantitatively by taking into account the measured value of the autocorrelation function (see Supporting Information) and, most importantly, the depolarization effects caused by the hyperfine interaction. Hudson et al.\cite{Hudson-2007} proposed a model for the phase evolution of the exciton-photon intermediate state of the XX-X decay cascade, which leads to Eq.~\ref{eq:fidelity_model} for the fidelity to the expected Bell state.
\begin{equation} \label{eq:fidelity_model}
f=\frac{1}{4}\left(1+\kappa g^{(1)}_{H,V}+\frac{2\kappa g^{(1)}_{H,V}}{1+\left(g^{(1)}_{H,V}S\tau_1/\hbar\right)^2}\right)
\end{equation}
In Eq.~\ref{eq:fidelity_model} $\kappa$ is the fraction of the photons generated from the QD exciton with respect to the background noise, $S$ is the FSS, $\tau_1$ is the radiative lifetime of the exciton, and $g^{(1)}_{H,V}$ is the first-order cross-coherence. Here we have assumed that cross dephasing is negligible and hence the first-order cross-coherence is given by $g^{(1)}_{H,V}=1/(1+\tau_1/\tau_{SS})$, where $\tau_{SS}$ is the characteristic time of spin scattering. The $\kappa$ coefficient can be inferred from autocorrelation measurements as $\kappa=1-g^{(2)}(0)$, with the measured value of exciton $g^{(2)}(0)$ equal to $0.03\pm0.01$.
Very recently, Huber et al. (see Supplementary Material of Ref.\citenum{Huber-2017}) have demonstrated that by using the typical spin scattering time from the literature\cite{Chekhovich-2013}, the fidelity closely followed the experimental results obtained for the droplet-etched QDs. Analogous conclusions were drawn in Ref.\citenum{Keil-2017}. Following a similar approach, we can calculate the expected fidelity distribution of our QDs obtained by droplet epitaxy, and we find that the large majority of the QDs, over 95\%, are potentially able to emit photon pairs with fidelity above the classical limit of 0.5.

\begin{figure*}[ht]
	\includegraphics[width=14cm]{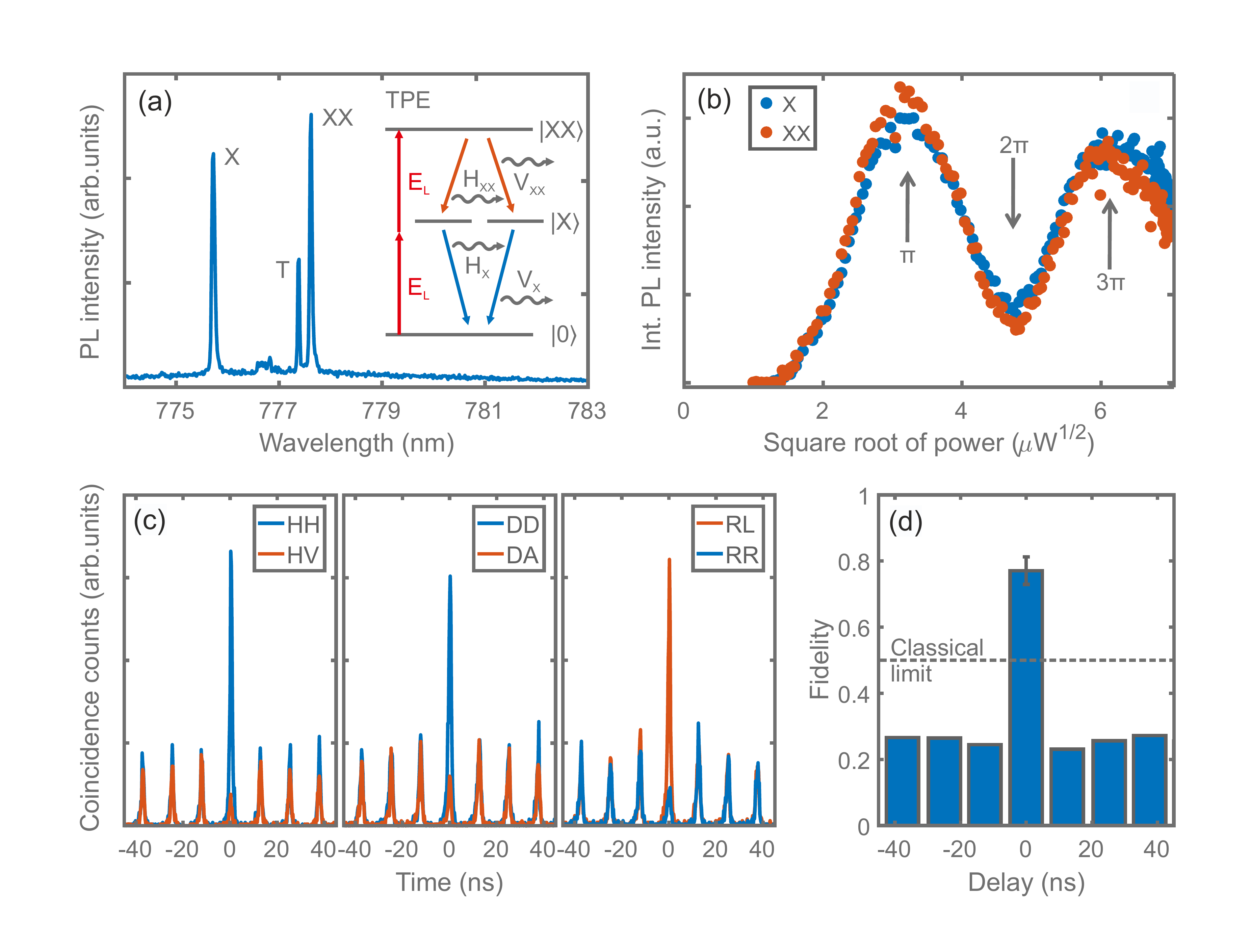}
	\caption{\label{fig:Entanglement} a) Emission spectrum of a single QD under two-photon resonant excitation (TPE). Inset describes the TPE process and the subsequent XX-X cascade. b) Integrated emission intensity of the exciton and biexciton lines under TPE as a function of the square root of the pump power. Rabi oscillations are highlighted. c) Cross-correlation measurements between X and XX emission for different polarization bases, namely linear, diagonal and circular. d) Fidelity to the maximally entangled state.}
\end{figure*}

Having illustrated that almost all our droplet epitaxy QDs have the potential to be used as entangled photon sources, we now address the characterization of the degree of entanglement of the emitted photons. Here, schemes for efficiently pumping the biexciton must be used because this transition is hardly visible in above-band excitation conditions (see Fig.~\ref{fig:microPL_eV}(a)). We employed a two-photon excitation resonant scheme in which the energy of the laser is tuned
half way between the exciton and biexciton recombination
energies\cite{Muller-2014}. Figure~\ref{fig:Entanglement}(a) shows how the emission energy spectrum is affected. While a contribution from a trion state is still present, the biexciton peak is as intense as the neutral exciton peak. The resonant character of the excitation process is demonstrated by the appearance of Rabi oscillations when the laser power is increased, as shown in Fig.~\ref{fig:Entanglement}(b).

To demonstrate the emission of polarization-entangled photons, we tuned the excitation power to the $\pi$ pulse. We considered a QD with the FSS of 2.6 $\pm$ 0.5 $\mu$eV, which is the value representative of a significant fraction of the QDs, and the radiative lifetime of 230 ps under resonant excitation. According to Eq.~\ref{eq:fidelity_model}, this should provide the entanglement fidelity of 0.77, above the classical limit. The cross-correlation measurements for this QD are shown in Fig.~\ref{fig:Entanglement}(c). Coincidences between the exciton and biexciton lines were counted in three different polarization bases, namely two pairs of orthogonal linear polarizations (H/V and D/A, where D is rotated by 45$^{\circ}$ with respect to H) in addition to the right (R) and left (L) handed circular polarization. The degree of correlation is calculated as $C_{AB}=(g_{XX,X}-g_{XX,\bar{X}})/(g_{XX,X}+g_{XX,\bar{X}})$, where $g_{XX,X}$ and $g_{XX,\bar{X}}$ are the coincidence counts between the exciton and biexciton emission, respectively for co-polarized and cross-polarized photons, integrated over the time window of a single pulse with the time bin of 6 ns. These data allow the direct estimation of the fidelity to the expected maximally entangled Bell state\cite{Hudson-2007} according to Eq.~\ref{eq:fidelity_coinc}.
\begin{equation} \label{eq:fidelity_coinc}
f=(1+C_{HV}+C_{DA}-C_{RL})/4
\end{equation}
The fidelity of the zero delay pulse is 0.77 $\pm$ 0.04 (error estimated with Gaussian propagation, assuming a Poisson distribution of the correlation counts), which is significantly above the upper limit for the classically correlated states (see Fig.~\ref{fig:Entanglement}(d)). The result is quantitatively consistent with the predictions of the previously discussed X states phase evolution model for GaAs QDs\cite{Huber-2017}, thus confirming that almost any QD in the ensemble is capable of delivering entangled photons with fidelities above the classical limit.

In conclusion, we have developed a novel class of droplet epitaxy QDs that can be used as a source of entangled photons in the spectral region of the $D_2$ lines of a cloud of Rb atoms.

The wide applicability of the DE growth scheme allowed us to choose the GaAs/AlGaAs materials system, and to show that 95\% of the emitters can deliver the photon pairs deterministically and with the fidelity to the expected Bell state that was above the classical limit.

To achieve this result, we addressed the main drawbacks of conventional DE by introducing droplet crystallization and barrier deposition at the high temperature close to the temperature used for high quality AlGaAs and GaAs deposition on (111)A. We have shown that this approach improves crystalline quality, leading to reduced spectral wandering and to reliable tuning of the emission wavelength due to limited interdiffusion during the capping process.

Since this technology is compatible with integration in optical microcavities for enhanced light extraction, and the growth can be easily adapted to different materials, it has the potential to become an ideal candidate for semiconductor-based sources of entangled photons.

\subsection{\label{sec:sub41}Samples fabrication}

The samples were grown in a Gen II molecular beam epitaxy system with an Arsenic valved cracker cell with the base pressure of $10^{-10}$ torr on intrinsic GaAs (111)A substrates.
After the oxide desorption, a 100 nm GaAs buffer layer was deposited at 520$^{\circ}$C with the Ga flux of 0.07 ML/s and the beam equivalent pressure of As of $3\times 10^{-5}$ torr. Then, an AlGaAs barrier with 50 nm of Al$_{0.3}$Ga$_{0.7}$As and 100 nm Al$_{0.15}$Ga$_{0.85}$As was grown with the total group-III fluxes of 0.1 and 0.082 ML/s, respectively. For both the buffer and barrier layers, the parameters for the temperature and III-V fluxes were chosen in order to minimize the formation of hillocks and provide flat AlGaAs surface prior to the QD deposition\cite{Esposito-2017}.

The Ga droplets were formed by depositing 0.4 MLs of Ga with the rate of 0.01 ML/s at 450$^{\circ}$C with the background pressure of less than $2\times 10^{-9}$ torr and then removing the sample from the chamber.						 
The GaAs QDs were formed by depositing 0.4 MLs of Ga at the rate of 0.01 ML/s at 450$^{\circ}$C with the background pressure of less than $2\times 10^{-9}$ torr and then supplying an As flux with the beam equivalent pressure of $3\times 10^{-5}$ torr at the substrate temperature of 500$^{\circ}$C. The QDs were then covered with a thin layer of Al$_{0.15}$Ga$_{0.85}$As grown at 500$^{\circ}$C followed by 100 nm Al$_{0.15}$Ga$_{0.85}$As, 50 nm of Al$_{0.3}$Ga$_{0.7}$As and a GaAs capping layer deposited at 520$^{\circ}$C. The QD capping procedure was not performed on the samples used for morphological characterization carried out by an AFM in the tapping mode using ultra sharp tips with the radius of 2 nm.

\subsection{Optical spectroscopy}

The sample was mounted inside a low-vibration continuous-flow helium cryostat working at 8 K. 

Under non-resonant excitation, single dot PL was excited by a 532 nm continuous wave laser at normal incidence through a 0.42 NA objective. 
A spatial filter, implemented with a single mode optical fiber, was occasionally added to the collection path to isolate single emitters.
The signal was sent to a double grating spectrometer, equipped with 1200 l/mm gratings, which let us achieve a 40 $\mu$eV spectral resolution in the 700-800 nm wavelength region, and finally acquired by a deep depletion, back-illuminated, LN2-CCD camera. 

Polarization-dependent spectra were acquired by adding a fixed linear polarizer and a rotating half-wave plate to the collection path. FSS was estimated as described in Ref.~\citenum{Abbarchi-2008}, resulting in the accuracy down to 1 $\mu$eV.

Michelson setup and details on the coherence time measurements are provided in Ref.~\citenum{Kammerer-2002}.

During the time-resolved experiments, the QDs were excited with a pulsed diode laser emitting at 440 nm and their signal was detected by a single photon avalanche detector with the time resolution of slightly above 50 ps.

Resonant two-photon excitation was accomplished using a Ti:sapphire femtosecond laser. The pulse duration was broadened from 100 fs to about 10 ps using a 4f pulse-shaper. Tunable notch filters with the bandwidth of 0.4 nm were placed in the collection path to suppress laser backscattering.

Photon correlation experiments were performed with a Hanbury-Brown-Twiss setup. The signal collected from the objective was sent to a non-polarizing beamsplitter and then to two polarization maintaining single mode fibers. The PL signal at the output of these fibers was sent to two independent spectrometers which could be tuned to direct a specific wavelength to an avalanche photodiode. The avalanche detectors were connected to the correlation electronics and each provided 500 ps timing jitter. For the fidelity measurements, exciton-biexciton cross-correlation was measured in different polarization bases, selected by linear polarizers and half- or quarter-wave plates inserted right after the beamsplitter.

\begin{acknowledgement}
	This work was financially supported by the European
	Research council (ERC) under the European Union's Horizon 2020 Research and Innovation Programme (SPQRel, Grant Agreement No. 679183).
	This  project  has  received  funding  from  the  European  Union's  Horizon  2020  research  and  innovation programme under the Marie Sk\l{}odowska-Curie grant agreement No. 721394.
	We thank Emanuele Grilli for fruitful discussions.
\end{acknowledgement}

\begin{suppinfo}
	
	Contour AFM map of a single QD. Ensemble photoluminescence. Lifetime and FSS data collected at different emission wavelengths. Typical polarization-resolved measurement and power-dependent analysis of single dot PL. Experimental autocorrelation function for X and XX emission.
	
\end{suppinfo}

\bibliography{bibliography}

\providecommand{\latin}[1]{#1}
\makeatletter
\providecommand{\doi}
  {\begingroup\let\do\@makeother\dospecials
  \catcode`\{=1 \catcode`\}=2 \doi@aux}
\providecommand{\doi@aux}[1]{\endgroup\texttt{#1}}
\makeatother
\providecommand*\mcitethebibliography{\thebibliography}
\csname @ifundefined\endcsname{endmcitethebibliography}
  {\let\endmcitethebibliography\endthebibliography}{}
\begin{mcitethebibliography}{66}
\providecommand*\natexlab[1]{#1}
\providecommand*\mciteSetBstSublistMode[1]{}
\providecommand*\mciteSetBstMaxWidthForm[2]{}
\providecommand*\mciteBstWouldAddEndPuncttrue
  {\def\EndOfBibitem{\unskip.}}
\providecommand*\mciteBstWouldAddEndPunctfalse
  {\let\EndOfBibitem\relax}
\providecommand*\mciteSetBstMidEndSepPunct[3]{}
\providecommand*\mciteSetBstSublistLabelBeginEnd[3]{}
\providecommand*\EndOfBibitem{}
\mciteSetBstSublistMode{f}
\mciteSetBstMaxWidthForm{subitem}{(\alph{mcitesubitemcount})}
\mciteSetBstSublistLabelBeginEnd
  {\mcitemaxwidthsubitemform\space}
  {\relax}
  {\relax}

\bibitem[Kimble(2008)]{Kimble-2008}
Kimble,~H.~J. The quantum internet. \emph{Nature} \textbf{2008}, \emph{453},
  1023--1030\relax
\mciteBstWouldAddEndPuncttrue
\mciteSetBstMidEndSepPunct{\mcitedefaultmidpunct}
{\mcitedefaultendpunct}{\mcitedefaultseppunct}\relax
\EndOfBibitem
\bibitem[Pan \latin{et~al.}(2012)Pan, Chen, Lu, Weinfurter, Zeilinger, and
  \ifmmode~\dot{Z}\else \.{Z}\fi{}ukowski]{Pan-2012}
Pan,~J.-W.; Chen,~Z.-B.; Lu,~C.-Y.; Weinfurter,~H.; Zeilinger,~A.;
  \ifmmode~\dot{Z}\else \.{Z}\fi{}ukowski,~M. Multiphoton entanglement and
  interferometry. \emph{Rev. Mod. Phys.} \textbf{2012}, \emph{84},
  777--838\relax
\mciteBstWouldAddEndPuncttrue
\mciteSetBstMidEndSepPunct{\mcitedefaultmidpunct}
{\mcitedefaultendpunct}{\mcitedefaultseppunct}\relax
\EndOfBibitem
\bibitem[Benson \latin{et~al.}(2000)Benson, Santori, Pelton, and
  Yamamoto]{Benson-2000}
Benson,~O.; Santori,~C.; Pelton,~M.; Yamamoto,~Y. Regulated and Entangled
  Photons from a Single Quantum Dot. \emph{Phys. Rev. Lett.} \textbf{2000},
  \emph{84}, 2513--2516\relax
\mciteBstWouldAddEndPuncttrue
\mciteSetBstMidEndSepPunct{\mcitedefaultmidpunct}
{\mcitedefaultendpunct}{\mcitedefaultseppunct}\relax
\EndOfBibitem
\bibitem[Lu and Pan(2014)Lu, and Pan]{Lu-2014}
Lu,~C.-Y.; Pan,~J.-W. Quantum optics: push-button photon entanglement.
  \emph{Nat. Photonics} \textbf{2014}, \emph{8}, 174--176\relax
\mciteBstWouldAddEndPuncttrue
\mciteSetBstMidEndSepPunct{\mcitedefaultmidpunct}
{\mcitedefaultendpunct}{\mcitedefaultseppunct}\relax
\EndOfBibitem
\bibitem[Bester \latin{et~al.}(2003)Bester, Nair, and Zunger]{Bester-2003}
Bester,~G.; Nair,~S.; Zunger,~A. Pseudopotential calculation of the excitonic
  fine structure of million-atom self-assembled
  ${\mathrm{In}}_{1-x}{\mathrm{Ga}}_{x}\mathrm{A}\mathrm{s}/\mathrm{G}\mathrm{a}\mathrm{A}\mathrm{s}$
  quantum dots. \emph{Phys. Rev. B} \textbf{2003}, \emph{67}, 161306\relax
\mciteBstWouldAddEndPuncttrue
\mciteSetBstMidEndSepPunct{\mcitedefaultmidpunct}
{\mcitedefaultendpunct}{\mcitedefaultseppunct}\relax
\EndOfBibitem
\bibitem[Seguin \latin{et~al.}(2005)Seguin, Schliwa, Rodt, P\"otschke, Pohl,
  and Bimberg]{Seguin-2005}
Seguin,~R.; Schliwa,~A.; Rodt,~S.; P\"otschke,~K.; Pohl,~U.~W.; Bimberg,~D.
  Size-Dependent Fine-Structure Splitting in Self-Organized
  $\mathrm{InAs}/\mathrm{GaAs}$ Quantum Dots. \emph{Phys. Rev. Lett.}
  \textbf{2005}, \emph{95}, 257402\relax
\mciteBstWouldAddEndPuncttrue
\mciteSetBstMidEndSepPunct{\mcitedefaultmidpunct}
{\mcitedefaultendpunct}{\mcitedefaultseppunct}\relax
\EndOfBibitem
\bibitem[Hudson \latin{et~al.}(2007)Hudson, Stevenson, Bennett, Young, Nicoll,
  Atkinson, Cooper, Ritchie, and Shields]{Hudson-2007}
Hudson,~A.~J.; Stevenson,~R.~M.; Bennett,~A.~J.; Young,~R.~J.; Nicoll,~C.~A.;
  Atkinson,~P.; Cooper,~K.; Ritchie,~D.~A.; Shields,~A.~J. Coherence of an
  Entangled Exciton-Photon State. \emph{Phys. Rev. Lett.} \textbf{2007},
  \emph{99}, 266802\relax
\mciteBstWouldAddEndPuncttrue
\mciteSetBstMidEndSepPunct{\mcitedefaultmidpunct}
{\mcitedefaultendpunct}{\mcitedefaultseppunct}\relax
\EndOfBibitem
\bibitem[Treu \latin{et~al.}(2012)Treu, Schneider, Huggenberger, Braun,
  Reitzenstein, H{\"o}fling, and Kamp]{Treu-2012}
Treu,~J.; Schneider,~C.; Huggenberger,~A.; Braun,~T.; Reitzenstein,~S.;
  H{\"o}fling,~S.; Kamp,~M. Substrate orientation dependent fine structure
  splitting of symmetric In(Ga)As/GaAs quantum dots. \emph{Appl. Phys. Lett.}
  \textbf{2012}, \emph{101}\relax
\mciteBstWouldAddEndPuncttrue
\mciteSetBstMidEndSepPunct{\mcitedefaultmidpunct}
{\mcitedefaultendpunct}{\mcitedefaultseppunct}\relax
\EndOfBibitem
\bibitem[Mano \latin{et~al.}(2010)Mano, Abbarchi, Kuroda, McSkimming, Ohtake,
  Mitsuishi, and Sakoda]{Mano-2010}
Mano,~T.; Abbarchi,~M.; Kuroda,~T.; McSkimming,~B.; Ohtake,~A.; Mitsuishi,~K.;
  Sakoda,~K. Self-Assembly of Symmetric GaAs Quantum Dots on (111)A Substrates:
  Suppression of Fine-Structure Splitting. \emph{Appl. Phys. Express}
  \textbf{2010}, \emph{3}, 065203\relax
\mciteBstWouldAddEndPuncttrue
\mciteSetBstMidEndSepPunct{\mcitedefaultmidpunct}
{\mcitedefaultendpunct}{\mcitedefaultseppunct}\relax
\EndOfBibitem
\bibitem[Huo \latin{et~al.}(2013)Huo, Rastelli, and Schmidt]{Huo-2013}
Huo,~Y.~H.; Rastelli,~A.; Schmidt,~O.~G. Ultra-small excitonic fine structure
  splitting in highly symmetric quantum dots on GaAs (001) substrate.
  \emph{Appl. Phys. Lett.} \textbf{2013}, \emph{102}, 152105\relax
\mciteBstWouldAddEndPuncttrue
\mciteSetBstMidEndSepPunct{\mcitedefaultmidpunct}
{\mcitedefaultendpunct}{\mcitedefaultseppunct}\relax
\EndOfBibitem
\bibitem[Yerino \latin{et~al.}(2014)Yerino, Simmonds, Liang, Jung, Schneider,
  Unsleber, Vo, Huffaker, H{\"o}fling, Kamp, and Lee]{Yerino-2014}
Yerino,~C.~D.; Simmonds,~P.~J.; Liang,~B.; Jung,~D.; Schneider,~C.;
  Unsleber,~S.; Vo,~M.; Huffaker,~D.~L.; H{\"o}fling,~S.; Kamp,~M.; Lee,~M.~L.
  Strain-driven growth of GaAs(111) quantum dots with low fine structure
  splitting. \emph{Appl. Phys. Lett.} \textbf{2014}, \emph{105}\relax
\mciteBstWouldAddEndPuncttrue
\mciteSetBstMidEndSepPunct{\mcitedefaultmidpunct}
{\mcitedefaultendpunct}{\mcitedefaultseppunct}\relax
\EndOfBibitem
\bibitem[Juska \latin{et~al.}(2013)Juska, Dimastrodonato, Mereni, Gocalinska,
  and Pelucchi]{Juska-2013}
Juska,~G.; Dimastrodonato,~V.; Mereni,~L.~O.; Gocalinska,~A.; Pelucchi,~E.
  Towards quantum-dot arrays of entangled photon emitters. \emph{Nat.
  Photonics} \textbf{2013}, \emph{7}, 527\relax
\mciteBstWouldAddEndPuncttrue
\mciteSetBstMidEndSepPunct{\mcitedefaultmidpunct}
{\mcitedefaultendpunct}{\mcitedefaultseppunct}\relax
\EndOfBibitem
\bibitem[Versteegh \latin{et~al.}(2014)Versteegh, Reimer, J\"ons, Dalacu,
  Poole, Gulinatti, Giudice, and Zwiller]{Versteegh-2014}
Versteegh,~M. A.~M.; Reimer,~M.~E.; J\"ons,~K.~D.; Dalacu,~D.; Poole,~P.~J.;
  Gulinatti,~A.; Giudice,~A.; Zwiller,~V. Observation of strongly entangled
  photon pairs from a nanowire quantum dot. \emph{Nat. Commun.} \textbf{2014},
  \emph{5}, 5298\relax
\mciteBstWouldAddEndPuncttrue
\mciteSetBstMidEndSepPunct{\mcitedefaultmidpunct}
{\mcitedefaultendpunct}{\mcitedefaultseppunct}\relax
\EndOfBibitem
\bibitem[Huber \latin{et~al.}(2017)Huber, Reindl, Huo, Huang, Wildmann,
  Schmidt, Rastelli, and Trotta]{Huber-2017}
Huber,~D.; Reindl,~M.; Huo,~Y.; Huang,~H.; Wildmann,~J.; Schmidt,~O.;
  Rastelli,~A.; Trotta,~R. Highly indistinguishable and strongly entangled
  photons from symmetric GaAs quantum dots. \emph{Nat. Commun.} \textbf{2017},
  \emph{8}, 15506\relax
\mciteBstWouldAddEndPuncttrue
\mciteSetBstMidEndSepPunct{\mcitedefaultmidpunct}
{\mcitedefaultendpunct}{\mcitedefaultseppunct}\relax
\EndOfBibitem
\bibitem[Sanguinetti and Koguchi(2013)Sanguinetti, and
  Koguchi]{Sanguinetti-2013}
Sanguinetti,~S.; Koguchi,~N. In \emph{Molecular Beam Epitaxy: From Research to
  Mass Production}; Henini,~M., Ed.; Elsevier: Oxford, 2013; p~95\relax
\mciteBstWouldAddEndPuncttrue
\mciteSetBstMidEndSepPunct{\mcitedefaultmidpunct}
{\mcitedefaultendpunct}{\mcitedefaultseppunct}\relax
\EndOfBibitem
\bibitem[Heyn \latin{et~al.}(2007)Heyn, Stemmann, Schramm, Welsch, Hansen, and
  Nemcsics]{Heyn-2007}
Heyn,~C.; Stemmann,~A.; Schramm,~A.; Welsch,~H.; Hansen,~W.; Nemcsics,~A.
  Regimes of GaAs quantum dot self-assembly by droplet epitaxy. \emph{Phys.
  Rev. B} \textbf{2007}, \emph{76}, 075317\relax
\mciteBstWouldAddEndPuncttrue
\mciteSetBstMidEndSepPunct{\mcitedefaultmidpunct}
{\mcitedefaultendpunct}{\mcitedefaultseppunct}\relax
\EndOfBibitem
\bibitem[Bietti \latin{et~al.}(2015)Bietti, Bocquel, Adorno, Mano, Keizer,
  Koenraad, and Sanguinetti]{Bietti-2015}
Bietti,~S.; Bocquel,~J.; Adorno,~S.; Mano,~T.; Keizer,~J.~G.; Koenraad,~P.~M.;
  Sanguinetti,~S. Precise shape engineering of epitaxial quantum dots by growth
  kinetics. \emph{Phys. Rev. B} \textbf{2015}, \emph{92}, 075425\relax
\mciteBstWouldAddEndPuncttrue
\mciteSetBstMidEndSepPunct{\mcitedefaultmidpunct}
{\mcitedefaultendpunct}{\mcitedefaultseppunct}\relax
\EndOfBibitem
\bibitem[Liu \latin{et~al.}(2014)Liu, Ha, Nakajima, Mano, Kuroda, Urbaszek,
  Kumano, Suemune, Sakuma, and Sakoda]{Liu-2014}
Liu,~X.; Ha,~N.; Nakajima,~H.; Mano,~T.; Kuroda,~T.; Urbaszek,~B.; Kumano,~H.;
  Suemune,~I.; Sakuma,~Y.; Sakoda,~K. Vanishing fine-structure splittings in
  telecommunication-wavelength quantum dots grown on (111)A surfaces by droplet
  epitaxy. \emph{Phys. Rev. B} \textbf{2014}, \emph{90}, 081301\relax
\mciteBstWouldAddEndPuncttrue
\mciteSetBstMidEndSepPunct{\mcitedefaultmidpunct}
{\mcitedefaultendpunct}{\mcitedefaultseppunct}\relax
\EndOfBibitem
\bibitem[Skiba-Szymanska \latin{et~al.}(2017)Skiba-Szymanska, Stevenson,
  Varnava, Felle, Huwer, M\"uller, Bennett, Lee, Farrer, Krysa, Spencer, Goff,
  Ritchie, Heffernan, and Shields]{Skiba-Szymanska-2017}
Skiba-Szymanska,~J.; Stevenson,~R.~M.; Varnava,~C.; Felle,~M.; Huwer,~J.;
  M\"uller,~T.; Bennett,~A.~J.; Lee,~J.~P.; Farrer,~I.; Krysa,~A.~B.;
  Spencer,~P.; Goff,~L.~E.; Ritchie,~D.~A.; Heffernan,~J.; Shields,~A.~J.
  Universal Growth Scheme for Quantum Dots with Low Fine-Structure Splitting at
  Various Emission Wavelengths. \emph{Phys. Rev. Applied} \textbf{2017},
  \emph{8}, 014013\relax
\mciteBstWouldAddEndPuncttrue
\mciteSetBstMidEndSepPunct{\mcitedefaultmidpunct}
{\mcitedefaultendpunct}{\mcitedefaultseppunct}\relax
\EndOfBibitem
\bibitem[Kuroda \latin{et~al.}(2013)Kuroda, Mano, Ha, Nakajima, Kumano,
  Urbaszek, Jo, Abbarchi, Sakuma, Sakoda, Suemune, Marie, and
  Amand]{Kuroda-2013}
Kuroda,~T.; Mano,~T.; Ha,~N.; Nakajima,~H.; Kumano,~H.; Urbaszek,~B.; Jo,~M.;
  Abbarchi,~M.; Sakuma,~Y.; Sakoda,~K.; Suemune,~I.; Marie,~X.; Amand,~T.
  Symmetric quantum dots as efficient sources of highly entangled photons:
  Violation of Bell's inequality without spectral and temporal filtering.
  \emph{Phys. Rev. B} \textbf{2013}, \emph{88}, 041306\relax
\mciteBstWouldAddEndPuncttrue
\mciteSetBstMidEndSepPunct{\mcitedefaultmidpunct}
{\mcitedefaultendpunct}{\mcitedefaultseppunct}\relax
\EndOfBibitem
\bibitem[Stevenson \latin{et~al.}(2006)Stevenson, Young, Atkinson, Cooper,
  Ritchie, and Shields]{Stevenson-2006}
Stevenson,~R.~M.; Young,~R.~J.; Atkinson,~P.; Cooper,~K.; Ritchie,~D.~A.;
  Shields,~A.~J. A semiconductor source of triggered entangled photon pairs.
  \emph{Nature} \textbf{2006}, \emph{439}, 179--182\relax
\mciteBstWouldAddEndPuncttrue
\mciteSetBstMidEndSepPunct{\mcitedefaultmidpunct}
{\mcitedefaultendpunct}{\mcitedefaultseppunct}\relax
\EndOfBibitem
\bibitem[Bennett \latin{et~al.}(2010)Bennett, Pooley, Stevenson, Ward, Patel,
  de~La~Giroday, Sk\"old, Farrer, Nicoll, Ritchie, and Shields]{Bennett-2010}
Bennett,~A.; Pooley,~M.; Stevenson,~R.; Ward,~M.; Patel,~R.;
  de~La~Giroday,~A.~B.; Sk\"old,~N.; Farrer,~I.; Nicoll,~C.; Ritchie,~D.;
  Shields,~A. Electric-field-induced coherent coupling of the exciton states in
  a single quantum dot. \emph{Nat. Phys.} \textbf{2010}, \emph{6},
  947--950\relax
\mciteBstWouldAddEndPuncttrue
\mciteSetBstMidEndSepPunct{\mcitedefaultmidpunct}
{\mcitedefaultendpunct}{\mcitedefaultseppunct}\relax
\EndOfBibitem
\bibitem[Trotta \latin{et~al.}(2016)Trotta, Mart{\'\i}n-S{\'a}nchez, Wildmann,
  Piredda, Reindl, Schimpf, Zallo, Stroj, Edlinger, and Rastelli]{Trotta-2016}
Trotta,~R.; Mart{\'\i}n-S{\'a}nchez,~J.; Wildmann,~J.~S.; Piredda,~G.;
  Reindl,~M.; Schimpf,~C.; Zallo,~E.; Stroj,~S.; Edlinger,~J.; Rastelli,~A.
  Wavelength-tunable sources of entangled photons interfaced with atomic
  vapours. \emph{Nat. Commun.} \textbf{2016}, \emph{7}, 10375\relax
\mciteBstWouldAddEndPuncttrue
\mciteSetBstMidEndSepPunct{\mcitedefaultmidpunct}
{\mcitedefaultendpunct}{\mcitedefaultseppunct}\relax
\EndOfBibitem
\bibitem[Muller \latin{et~al.}(2009)Muller, Fang, Lawall, and
  Solomon]{Muller-2009}
Muller,~A.; Fang,~W.; Lawall,~J.; Solomon,~G.~S. Creating
  Polarization-Entangled Photon Pairs from a Semiconductor Quantum Dot Using
  the Optical Stark Effect. \emph{Phys. Rev. Lett.} \textbf{2009}, \emph{103},
  217402\relax
\mciteBstWouldAddEndPuncttrue
\mciteSetBstMidEndSepPunct{\mcitedefaultmidpunct}
{\mcitedefaultendpunct}{\mcitedefaultseppunct}\relax
\EndOfBibitem
\bibitem[Singh and Bester(2009)Singh, and Bester]{Singh-2009}
Singh,~R.; Bester,~G. Nanowire Quantum Dots as an Ideal Source of Entangled
  Photon Pairs. \emph{Phys. Rev. Lett.} \textbf{2009}, \emph{103}, 063601\relax
\mciteBstWouldAddEndPuncttrue
\mciteSetBstMidEndSepPunct{\mcitedefaultmidpunct}
{\mcitedefaultendpunct}{\mcitedefaultseppunct}\relax
\EndOfBibitem
\bibitem[Schliwa \latin{et~al.}(2009)Schliwa, Winkelnkemper, Lochmann, Stock,
  and Bimberg]{Schliwa-2009}
Schliwa,~A.; Winkelnkemper,~M.; Lochmann,~A.; Stock,~E.; Bimberg,~D.
  In(Ga)As/GaAs quantum dots grown on a (111) surface as ideal sources of
  entangled photon pairs. \emph{Phys. Rev. B} \textbf{2009}, \emph{80},
  161307\relax
\mciteBstWouldAddEndPuncttrue
\mciteSetBstMidEndSepPunct{\mcitedefaultmidpunct}
{\mcitedefaultendpunct}{\mcitedefaultseppunct}\relax
\EndOfBibitem
\bibitem[Akopian \latin{et~al.}(2011)Akopian, Wang, Rastelli, Schmidt, and
  Zwiller]{Akopian-2011}
Akopian,~N.; Wang,~L.; Rastelli,~A.; Schmidt,~O.; Zwiller,~V. Hybrid
  semiconductor-atomic interface: slowing down single photons from a quantum
  dot. \emph{Nat. Photonics} \textbf{2011}, \emph{5}, 230--233\relax
\mciteBstWouldAddEndPuncttrue
\mciteSetBstMidEndSepPunct{\mcitedefaultmidpunct}
{\mcitedefaultendpunct}{\mcitedefaultseppunct}\relax
\EndOfBibitem
\bibitem[Huang \latin{et~al.}(2017)Huang, Trotta, Huo, Lettner, Wildmann,
  Mart{\'\i}n-S{\'a}nchez, Huber, Reindl, Zhang, Zallo, Schmidt, and
  Rastelli]{Huang-2017}
Huang,~H.; Trotta,~R.; Huo,~Y.; Lettner,~T.; Wildmann,~J.~S.;
  Mart{\'\i}n-S{\'a}nchez,~J.; Huber,~D.; Reindl,~M.; Zhang,~J.; Zallo,~E.;
  Schmidt,~O.~G.; Rastelli,~A. Electrically-pumped wavelength-tunable GaAs
  quantum dots interfaced with rubidium atoms. \emph{ACS photonics}
  \textbf{2017}, \emph{4}, 868--872\relax
\mciteBstWouldAddEndPuncttrue
\mciteSetBstMidEndSepPunct{\mcitedefaultmidpunct}
{\mcitedefaultendpunct}{\mcitedefaultseppunct}\relax
\EndOfBibitem
\bibitem[Patel \latin{et~al.}(2010)Patel, Bennett, Farrer, Nicoll, Ritchie, and
  Shields]{Patel-2010}
Patel,~R.~B.; Bennett,~A.~J.; Farrer,~I.; Nicoll,~C.~A.; Ritchie,~D.~A.;
  Shields,~A.~J. Two-photon interference of the emission from electrically
  tunable remote quantum dots. \emph{Nat. Photonics} \textbf{2010}, \emph{4},
  632--635\relax
\mciteBstWouldAddEndPuncttrue
\mciteSetBstMidEndSepPunct{\mcitedefaultmidpunct}
{\mcitedefaultendpunct}{\mcitedefaultseppunct}\relax
\EndOfBibitem
\bibitem[Mantovani \latin{et~al.}(2004)Mantovani, Sanguinetti, Guzzi, Grilli,
  Gurioli, Watanabe, and Koguchi]{Mantovani-2004}
Mantovani,~V.; Sanguinetti,~S.; Guzzi,~M.; Grilli,~E.; Gurioli,~M.;
  Watanabe,~K.; Koguchi,~N. Low density GaAs/AlGaAs quantum dots grown by
  modified droplet epitaxy. \emph{J. Appl. Phys.} \textbf{2004}, \emph{96},
  4416--4420\relax
\mciteBstWouldAddEndPuncttrue
\mciteSetBstMidEndSepPunct{\mcitedefaultmidpunct}
{\mcitedefaultendpunct}{\mcitedefaultseppunct}\relax
\EndOfBibitem
\bibitem[Mano \latin{et~al.}(2009)Mano, Abbarchi, Kuroda, Mastrandrea,
  Vinattieri, Sanguinetti, Sakoda, and Gurioli]{Mano-2009}
Mano,~T.; Abbarchi,~M.; Kuroda,~T.; Mastrandrea,~C.; Vinattieri,~A.;
  Sanguinetti,~S.; Sakoda,~K.; Gurioli,~M. Ultra-narrow emission from single
  GaAs self-assembled quantum dots grown by droplet epitaxy.
  \emph{Nanotechnology} \textbf{2009}, \emph{20}, 395601\relax
\mciteBstWouldAddEndPuncttrue
\mciteSetBstMidEndSepPunct{\mcitedefaultmidpunct}
{\mcitedefaultendpunct}{\mcitedefaultseppunct}\relax
\EndOfBibitem
\bibitem[Jo \latin{et~al.}(2012)Jo, Mano, Abbarchi, Kuroda, Sakuma, and
  Sakoda]{Jo-2012}
Jo,~M.; Mano,~T.; Abbarchi,~M.; Kuroda,~T.; Sakuma,~Y.; Sakoda,~K.
  Self-Limiting Growth of Hexagonal and Triangular Quantum Dots on (111)A.
  \emph{Cryst. Growth Des.} \textbf{2012}, \emph{12}, 1411--1415\relax
\mciteBstWouldAddEndPuncttrue
\mciteSetBstMidEndSepPunct{\mcitedefaultmidpunct}
{\mcitedefaultendpunct}{\mcitedefaultseppunct}\relax
\EndOfBibitem
\bibitem[Watanabe \latin{et~al.}(2001)Watanabe, Tsukamoto, Gotoh, and
  Koguchi]{Watanabe-2001}
Watanabe,~K.; Tsukamoto,~S.; Gotoh,~Y.; Koguchi,~N. Photoluminescence studies
  of GaAs quantum dots grown by droplet epitaxy. \emph{J. Cryst. Growth}
  \textbf{2001}, \emph{227}, 1073--1077\relax
\mciteBstWouldAddEndPuncttrue
\mciteSetBstMidEndSepPunct{\mcitedefaultmidpunct}
{\mcitedefaultendpunct}{\mcitedefaultseppunct}\relax
\EndOfBibitem
\bibitem[Bietti \latin{et~al.}(2013)Bietti, Somaschini, and
  Sanguinetti]{Bietti-2013}
Bietti,~S.; Somaschini,~C.; Sanguinetti,~S. Crystallization kinetics of Ga
  metallic nano-droplets under As flux. \emph{Nanotechnology} \textbf{2013},
  \emph{24}, 205603\relax
\mciteBstWouldAddEndPuncttrue
\mciteSetBstMidEndSepPunct{\mcitedefaultmidpunct}
{\mcitedefaultendpunct}{\mcitedefaultseppunct}\relax
\EndOfBibitem
\bibitem[Reyes \latin{et~al.}(2013)Reyes, Smereka, Nothern, Millunchick,
  Bietti, Somaschini, Sanguinetti, and Frigeri]{Reyes-2013}
Reyes,~K.; Smereka,~P.; Nothern,~D.; Millunchick,~J.~M.; Bietti,~S.;
  Somaschini,~C.; Sanguinetti,~S.; Frigeri,~C. Unified model of droplet epitaxy
  for compound semiconductor nanostructures: Experiments and theory.
  \emph{Phys. Rev. B} \textbf{2013}, \emph{87}, 165406\relax
\mciteBstWouldAddEndPuncttrue
\mciteSetBstMidEndSepPunct{\mcitedefaultmidpunct}
{\mcitedefaultendpunct}{\mcitedefaultseppunct}\relax
\EndOfBibitem
\bibitem[Somaschini \latin{et~al.}(2009)Somaschini, Bietti, Koguchi, and
  Sanguinetti]{Somaschini-2009}
Somaschini,~C.; Bietti,~S.; Koguchi,~N.; Sanguinetti,~S. Fabrication of
  Multiple Concentric Nanoring Structures. \emph{Nano Letters} \textbf{2009},
  \emph{9}, 3419--3424\relax
\mciteBstWouldAddEndPuncttrue
\mciteSetBstMidEndSepPunct{\mcitedefaultmidpunct}
{\mcitedefaultendpunct}{\mcitedefaultseppunct}\relax
\EndOfBibitem
\bibitem[Sato \latin{et~al.}(1994)Sato, Fahy, and Joyce]{Sato-1994}
Sato,~K.; Fahy,~M.; Joyce,~B. Reflection high energy electron diffraction
  intensity oscillation study of the growth of GaAs on GaAs(111)A. \emph{Surf.
  Sci.} \textbf{1994}, \emph{315}, 105 -- 111\relax
\mciteBstWouldAddEndPuncttrue
\mciteSetBstMidEndSepPunct{\mcitedefaultmidpunct}
{\mcitedefaultendpunct}{\mcitedefaultseppunct}\relax
\EndOfBibitem
\bibitem[Bietti \latin{et~al.}(2014)Bietti, Somaschini, Esposito, Fedorov, and
  Sanguinetti]{Bietti-2014}
Bietti,~S.; Somaschini,~C.; Esposito,~L.; Fedorov,~A.; Sanguinetti,~S. Gallium
  surface diffusion on GaAs (001) surfaces measured by crystallization dynamics
  of Ga droplets. \emph{J. Appl. Phys.} \textbf{2014}, \emph{116}, 114311\relax
\mciteBstWouldAddEndPuncttrue
\mciteSetBstMidEndSepPunct{\mcitedefaultmidpunct}
{\mcitedefaultendpunct}{\mcitedefaultseppunct}\relax
\EndOfBibitem
\bibitem[Keizer \latin{et~al.}(2010)Keizer, Bocquel, Koenraad, Mano, Noda, and
  Sakoda]{Keizer-2010}
Keizer,~J.~G.; Bocquel,~J.; Koenraad,~P.~M.; Mano,~T.; Noda,~T.; Sakoda,~K.
  Atomic scale analysis of self assembled GaAs/AlGaAs quantum dots grown by
  droplet epitaxy. \emph{Appl. Phys. Lett.} \textbf{2010}, \emph{96},
  062101\relax
\mciteBstWouldAddEndPuncttrue
\mciteSetBstMidEndSepPunct{\mcitedefaultmidpunct}
{\mcitedefaultendpunct}{\mcitedefaultseppunct}\relax
\EndOfBibitem
\bibitem[Ha \latin{et~al.}(2015)Ha, Mano, Chou, Wu, Cheng, Bocquel, Koenraad,
  Ohtake, Sakuma, Sakoda, and Kuroda]{Ha-2015}
Ha,~N.; Mano,~T.; Chou,~Y.-L.; Wu,~Y.-N.; Cheng,~S.-J.; Bocquel,~J.;
  Koenraad,~P.~M.; Ohtake,~A.; Sakuma,~Y.; Sakoda,~K.; Kuroda,~T.
  Size-dependent line broadening in the emission spectra of single GaAs quantum
  dots: Impact of surface charge on spectral diffusion. \emph{Phys. Rev. B}
  \textbf{2015}, \emph{92}, 075306\relax
\mciteBstWouldAddEndPuncttrue
\mciteSetBstMidEndSepPunct{\mcitedefaultmidpunct}
{\mcitedefaultendpunct}{\mcitedefaultseppunct}\relax
\EndOfBibitem
\bibitem[Schlesinger and Kuech(1986)Schlesinger, and Kuech]{Schlesinger-1986}
Schlesinger,~T.~E.; Kuech,~T. Determination of the interdiffusion of Al and Ga
  in undoped (Al,Ga)As/GaAs quantum wells. \emph{Appl. Phys. Lett.}
  \textbf{1986}, \emph{49}, 519--521\relax
\mciteBstWouldAddEndPuncttrue
\mciteSetBstMidEndSepPunct{\mcitedefaultmidpunct}
{\mcitedefaultendpunct}{\mcitedefaultseppunct}\relax
\EndOfBibitem
\bibitem[Bracht \latin{et~al.}(1999)Bracht, Haller, Eberl, and
  Cardona]{Bracht-1999}
Bracht,~H.; Haller,~E.~E.; Eberl,~K.; Cardona,~M. Self- and interdiffusion in
  AlXGa1-XAs/GaAs isotope heterostructures. \emph{Appl. Phys. Lett.}
  \textbf{1999}, \emph{74}, 49--51\relax
\mciteBstWouldAddEndPuncttrue
\mciteSetBstMidEndSepPunct{\mcitedefaultmidpunct}
{\mcitedefaultendpunct}{\mcitedefaultseppunct}\relax
\EndOfBibitem
\bibitem[Abbarchi \latin{et~al.}(2008)Abbarchi, Mastrandrea, Kuroda, Mano,
  Sakoda, Koguchi, Sanguinetti, Vinattieri, and Gurioli]{Abbarchi-2008}
Abbarchi,~M.; Mastrandrea,~C.~A.; Kuroda,~T.; Mano,~T.; Sakoda,~K.;
  Koguchi,~N.; Sanguinetti,~S.; Vinattieri,~A.; Gurioli,~M. Exciton fine
  structure in strain-free
  $\text{GaAs}/{\text{Al}}_{0.3}{\text{Ga}}_{0.7}\text{As}$ quantum dots:
  Extrinsic effects. \emph{Phys. Rev. B} \textbf{2008}, \emph{78}, 125321\relax
\mciteBstWouldAddEndPuncttrue
\mciteSetBstMidEndSepPunct{\mcitedefaultmidpunct}
{\mcitedefaultendpunct}{\mcitedefaultseppunct}\relax
\EndOfBibitem
\bibitem[Luo and Zunger(2011)Luo, and Zunger]{Luo-2011}
Luo,~J.-W.; Zunger,~A. Geometry of epitaxial GaAs/(Al,Ga)As quantum dots as
  seen by excitonic spectroscopy. \emph{Phys. Rev. B} \textbf{2011}, \emph{84},
  235317\relax
\mciteBstWouldAddEndPuncttrue
\mciteSetBstMidEndSepPunct{\mcitedefaultmidpunct}
{\mcitedefaultendpunct}{\mcitedefaultseppunct}\relax
\EndOfBibitem
\bibitem[Juska \latin{et~al.}(2015)Juska, Murray, Dimastrodonato, Chung,
  Moroni, Gocalinska, and Pelucchi]{Juska-2015}
Juska,~G.; Murray,~E.; Dimastrodonato,~V.; Chung,~T.~H.; Moroni,~S.~T.;
  Gocalinska,~A.; Pelucchi,~E. Conditions for entangled photon emission from
  (111)B site-controlled pyramidal quantum dots. \emph{J. Appl. Phys.}
  \textbf{2015}, \emph{117}, 134302\relax
\mciteBstWouldAddEndPuncttrue
\mciteSetBstMidEndSepPunct{\mcitedefaultmidpunct}
{\mcitedefaultendpunct}{\mcitedefaultseppunct}\relax
\EndOfBibitem
\bibitem[Jahn \latin{et~al.}(2015)Jahn, Munsch, B\'eguin, Kuhlmann, Renggli,
  Huo, Ding, Trotta, Reindl, Schmidt, Rastelli, Treutlein, and
  Warburton]{Jahn-2015}
Jahn,~J.-P.; Munsch,~M.; B\'eguin,~L.; Kuhlmann,~A.~V.; Renggli,~M.; Huo,~Y.;
  Ding,~F.; Trotta,~R.; Reindl,~M.; Schmidt,~O.~G.; Rastelli,~A.;
  Treutlein,~P.; Warburton,~R.~J. An artificial Rb atom in a semiconductor with
  lifetime-limited linewidth. \emph{Phys. Rev. B} \textbf{2015}, \emph{92},
  245439\relax
\mciteBstWouldAddEndPuncttrue
\mciteSetBstMidEndSepPunct{\mcitedefaultmidpunct}
{\mcitedefaultendpunct}{\mcitedefaultseppunct}\relax
\EndOfBibitem
\bibitem[Uskov \latin{et~al.}(2001)Uskov, Magnusdottir, Tromborg, M{\o}rk, and
  Lang]{Uskov-2001}
Uskov,~A.~V.; Magnusdottir,~I.; Tromborg,~B.; M{\o}rk,~J.; Lang,~R. Line
  broadening caused by Coulomb carrier–carrier correlations and dynamics of
  carrier capture and emission in quantum dots. \emph{Appl. Phys. Lett.}
  \textbf{2001}, \emph{79}, 1679--1681\relax
\mciteBstWouldAddEndPuncttrue
\mciteSetBstMidEndSepPunct{\mcitedefaultmidpunct}
{\mcitedefaultendpunct}{\mcitedefaultseppunct}\relax
\EndOfBibitem
\bibitem[Berthelot \latin{et~al.}(2006)Berthelot, Favero, Cassabois, Voisin,
  Delalande, Roussignol, Ferreira, and G{\'e}rard]{Berthelot-2006}
Berthelot,~A.; Favero,~I.; Cassabois,~G.; Voisin,~C.; Delalande,~C.;
  Roussignol,~P.; Ferreira,~R.; G{\'e}rard,~J.-M. Unconventional motional
  narrowing in the optical spectrum of a semiconductor quantum dot. \emph{Nat.
  Phys.} \textbf{2006}, \emph{2}, 759--764\relax
\mciteBstWouldAddEndPuncttrue
\mciteSetBstMidEndSepPunct{\mcitedefaultmidpunct}
{\mcitedefaultendpunct}{\mcitedefaultseppunct}\relax
\EndOfBibitem
\bibitem[Kuhlmann \latin{et~al.}(2013)Kuhlmann, Houel, Ludwig, Greuter, Reuter,
  Wieck, Poggio, and Warburton]{Kuhlmann-2013}
Kuhlmann,~A.~V.; Houel,~J.; Ludwig,~A.; Greuter,~L.; Reuter,~D.; Wieck,~A.~D.;
  Poggio,~M.; Warburton,~R.~J. Charge noise and spin noise in a semiconductor
  quantum device. \emph{Nat. Phys.} \textbf{2013}, \emph{9}, 570--575\relax
\mciteBstWouldAddEndPuncttrue
\mciteSetBstMidEndSepPunct{\mcitedefaultmidpunct}
{\mcitedefaultendpunct}{\mcitedefaultseppunct}\relax
\EndOfBibitem
\bibitem[Kammerer \latin{et~al.}(2002)Kammerer, Cassabois, Voisin, Perrin,
  Delalande, Roussignol, and G{\'{e}}rard]{Kammerer-2002}
Kammerer,~C.; Cassabois,~G.; Voisin,~C.; Perrin,~M.; Delalande,~C.;
  Roussignol,~P.; G{\'{e}}rard,~J.~M. Interferometric correlation spectroscopy
  in single quantum dots. \emph{Appl. Phys. Lett.} \textbf{2002}, \emph{81},
  2737--2739\relax
\mciteBstWouldAddEndPuncttrue
\mciteSetBstMidEndSepPunct{\mcitedefaultmidpunct}
{\mcitedefaultendpunct}{\mcitedefaultseppunct}\relax
\EndOfBibitem
\bibitem[Abbarchi \latin{et~al.}(2011)Abbarchi, Kuroda, Duval, Mano, and
  Sakoda]{Abbarchi-2011}
Abbarchi,~M.; Kuroda,~T.; Duval,~R.; Mano,~T.; Sakoda,~K. Scanning Fabry-Pérot
  interferometer with largely tuneable free spectral range for high resolution
  spectroscopy of single quantum dots. \emph{Rev. Sci. Instrum.} \textbf{2011},
  \emph{82}, 073103\relax
\mciteBstWouldAddEndPuncttrue
\mciteSetBstMidEndSepPunct{\mcitedefaultmidpunct}
{\mcitedefaultendpunct}{\mcitedefaultseppunct}\relax
\EndOfBibitem
\bibitem[Bayer \latin{et~al.}(2002)Bayer, Ortner, Stern, Kuther, Gorbunov,
  Forchel, Hawrylak, Fafard, Hinzer, Reinecke, Walck, Reithmaier, Klopf, and
  Sch\"afer]{Bayer-2002}
Bayer,~M.; Ortner,~G.; Stern,~O.; Kuther,~A.; Gorbunov,~A.~A.; Forchel,~A.;
  Hawrylak,~P.; Fafard,~S.; Hinzer,~K.; Reinecke,~T.~L.; Walck,~S.~N.;
  Reithmaier,~J.~P.; Klopf,~F.; Sch\"afer,~F. Fine structure of neutral and
  charged excitons in self-assembled In(Ga)As/(Al)GaAs quantum dots.
  \emph{Phys. Rev. B} \textbf{2002}, \emph{65}, 195315\relax
\mciteBstWouldAddEndPuncttrue
\mciteSetBstMidEndSepPunct{\mcitedefaultmidpunct}
{\mcitedefaultendpunct}{\mcitedefaultseppunct}\relax
\EndOfBibitem
\bibitem[Ohtake \latin{et~al.}(2015)Ohtake, Ha, and Mano]{Ohtake-2015}
Ohtake,~A.; Ha,~N.; Mano,~T. Extremely High- and Low-Density of Ga Droplets on
  GaAs{111}A,B: Surface-Polarity Dependence. \emph{Cryst. Growth Des.}
  \textbf{2015}, \emph{15}, 485--488\relax
\mciteBstWouldAddEndPuncttrue
\mciteSetBstMidEndSepPunct{\mcitedefaultmidpunct}
{\mcitedefaultendpunct}{\mcitedefaultseppunct}\relax
\EndOfBibitem
\bibitem[Huo \latin{et~al.}(2014)Huo, K{\v{r}}{\'{a}}pek, Rastelli, and
  Schmidt]{Huo-2014}
Huo,~Y.~H.; K{\v{r}}{\'{a}}pek,~V.; Rastelli,~A.; Schmidt,~O.~G. {Volume
  dependence of excitonic fine structure splitting in geometrically similar
  quantum dots}. \emph{Phys. Rev. B} \textbf{2014}, \emph{90}, 041304\relax
\mciteBstWouldAddEndPuncttrue
\mciteSetBstMidEndSepPunct{\mcitedefaultmidpunct}
{\mcitedefaultendpunct}{\mcitedefaultseppunct}\relax
\EndOfBibitem
\bibitem[Young \latin{et~al.}(2005)Young, Stevenson, Shields, Atkinson, Cooper,
  Ritchie, Groom, Tartakovskii, and Skolnick]{Young-2005}
Young,~R.~J.; Stevenson,~R.~M.; Shields,~A.~J.; Atkinson,~P.; Cooper,~K.;
  Ritchie,~D.~A.; Groom,~K.~M.; Tartakovskii,~A.~I.; Skolnick,~M.~S. Inversion
  of exciton level splitting in quantum dots. \emph{Phys. Rev. B}
  \textbf{2005}, \emph{72}, 113305\relax
\mciteBstWouldAddEndPuncttrue
\mciteSetBstMidEndSepPunct{\mcitedefaultmidpunct}
{\mcitedefaultendpunct}{\mcitedefaultseppunct}\relax
\EndOfBibitem
\bibitem[Keil \latin{et~al.}(2017)Keil, Zopf, Chen, H{\"o}fer, Zhang, Ding, and
  Schmidt]{Keil-2017}
Keil,~R.; Zopf,~M.; Chen,~Y.; H{\"o}fer,~B.; Zhang,~J.; Ding,~F.;
  Schmidt,~O.~G. Solid-state ensemble of highly entangled photon sources at
  rubidium atomic transitions. \emph{Nat. Commun.} \textbf{2017}, \emph{8},
  15501\relax
\mciteBstWouldAddEndPuncttrue
\mciteSetBstMidEndSepPunct{\mcitedefaultmidpunct}
{\mcitedefaultendpunct}{\mcitedefaultseppunct}\relax
\EndOfBibitem
\bibitem[Raymond \latin{et~al.}(1996)Raymond, Fafard, Poole, Wojs, Hawrylak,
  Charbonneau, Leonard, Leon, Petroff, and Merz]{Raymond-1996}
Raymond,~S.; Fafard,~S.; Poole,~P.~J.; Wojs,~A.; Hawrylak,~P.; Charbonneau,~S.;
  Leonard,~D.; Leon,~R.; Petroff,~P.~M.; Merz,~J.~L. State filling and
  time-resolved photoluminescence of excited states in
  ${\mathrm{In}}_{\mathit{x}}$${\mathrm{Ga}}_{1\mathrm{\ensuremath{-}}\mathit{x}}$As/GaAs
  self-assembled quantum dots. \emph{Phys. Rev. B} \textbf{1996}, \emph{54},
  11548--11554\relax
\mciteBstWouldAddEndPuncttrue
\mciteSetBstMidEndSepPunct{\mcitedefaultmidpunct}
{\mcitedefaultendpunct}{\mcitedefaultseppunct}\relax
\EndOfBibitem
\bibitem[Tighineanu \latin{et~al.}(2013)Tighineanu, Daveau, Lee, Song, Stobbe,
  and Lodahl]{Tighineanu-2013}
Tighineanu,~P.; Daveau,~R.; Lee,~E.~H.; Song,~J.~D.; Stobbe,~S.; Lodahl,~P.
  Decay dynamics and exciton localization in large GaAs quantum dots grown by
  droplet epitaxy. \emph{Phys. Rev. B} \textbf{2013}, \emph{88}, 155320\relax
\mciteBstWouldAddEndPuncttrue
\mciteSetBstMidEndSepPunct{\mcitedefaultmidpunct}
{\mcitedefaultendpunct}{\mcitedefaultseppunct}\relax
\EndOfBibitem
\bibitem[Dalgarno \latin{et~al.}(2008)Dalgarno, Smith, McFarlane, Gerardot,
  Karrai, Badolato, Petroff, and Warburton]{Dalgarno-2008}
Dalgarno,~P.~A.; Smith,~J.~M.; McFarlane,~J.; Gerardot,~B.~D.; Karrai,~K.;
  Badolato,~A.; Petroff,~P.~M.; Warburton,~R.~J. Coulomb interactions in single
  charged self-assembled quantum dots: Radiative lifetime and recombination
  energy. \emph{Phys. Rev. B} \textbf{2008}, \emph{77}, 245311\relax
\mciteBstWouldAddEndPuncttrue
\mciteSetBstMidEndSepPunct{\mcitedefaultmidpunct}
{\mcitedefaultendpunct}{\mcitedefaultseppunct}\relax
\EndOfBibitem
\bibitem[Langbein \latin{et~al.}(2004)Langbein, Borri, Woggon, Stavarache,
  Reuter, and Wieck]{Langbein-2004}
Langbein,~W.; Borri,~P.; Woggon,~U.; Stavarache,~V.; Reuter,~D.; Wieck,~A.~D.
  Radiatively limited dephasing in InAs quantum dots. \emph{Phys. Rev. B}
  \textbf{2004}, \emph{70}, 033301\relax
\mciteBstWouldAddEndPuncttrue
\mciteSetBstMidEndSepPunct{\mcitedefaultmidpunct}
{\mcitedefaultendpunct}{\mcitedefaultseppunct}\relax
\EndOfBibitem
\bibitem[Trotta \latin{et~al.}(2014)Trotta, Wildmann, Zallo, Schmidt, and
  Rastelli]{Trotta-2015-NL}
Trotta,~R.; Wildmann,~J.~S.; Zallo,~E.; Schmidt,~O.~G.; Rastelli,~A. Highly
  Entangled Photons from Hybrid Piezoelectric-Semiconductor Quantum Dot
  Devices. \emph{Nano Lett.} \textbf{2014}, \emph{14}, 3439--3444\relax
\mciteBstWouldAddEndPuncttrue
\mciteSetBstMidEndSepPunct{\mcitedefaultmidpunct}
{\mcitedefaultendpunct}{\mcitedefaultseppunct}\relax
\EndOfBibitem
\bibitem[{Akopian} \latin{et~al.}(2013){Akopian}, {Trotta}, {Zallo}, {Kumar},
  {Atkinson}, {Rastelli}, {Schmidt}, and {Zwiller}]{Akopian-2013}
{Akopian},~N.; {Trotta},~R.; {Zallo},~E.; {Kumar},~S.; {Atkinson},~P.;
  {Rastelli},~A.; {Schmidt},~O.~G.; {Zwiller},~V. {An artificial atom locked to
  natural atoms}. \emph{arXiv.org e-Print archive.} \textbf{2013},
  arXiv:1302.2005\relax
\mciteBstWouldAddEndPuncttrue
\mciteSetBstMidEndSepPunct{\mcitedefaultmidpunct}
{\mcitedefaultendpunct}{\mcitedefaultseppunct}\relax
\EndOfBibitem
\bibitem[Chekhovich \latin{et~al.}(2013)Chekhovich, Makhonin, Tartakovskii,
  Yacoby, Bluhm, Nowack, and Vandersypen]{Chekhovich-2013}
Chekhovich,~E.; Makhonin,~M.; Tartakovskii,~A.; Yacoby,~A.; Bluhm,~H.;
  Nowack,~K.; Vandersypen,~L. Nuclear spin effects in semiconductor quantum
  dots. \emph{Nat. Mater.} \textbf{2013}, \emph{12}, 494--504\relax
\mciteBstWouldAddEndPuncttrue
\mciteSetBstMidEndSepPunct{\mcitedefaultmidpunct}
{\mcitedefaultendpunct}{\mcitedefaultseppunct}\relax
\EndOfBibitem
\bibitem[M{\"u}ller \latin{et~al.}(2014)M{\"u}ller, Bounouar, J{\"o}ns,
  Gl{\"a}ssl, and Michler]{Muller-2014}
M{\"u}ller,~M.; Bounouar,~S.; J{\"o}ns,~K.~D.; Gl{\"a}ssl,~M.; Michler,~P.
  On-demand generation of indistinguishable polarization-entangled photon
  pairs. \emph{Nat. Photonics} \textbf{2014}, \emph{8}, 224--228\relax
\mciteBstWouldAddEndPuncttrue
\mciteSetBstMidEndSepPunct{\mcitedefaultmidpunct}
{\mcitedefaultendpunct}{\mcitedefaultseppunct}\relax
\EndOfBibitem
\bibitem[Esposito \latin{et~al.}(2017)Esposito, Bietti, Fedorov, N\"otzel, and
  Sanguinetti]{Esposito-2017}
Esposito,~L.; Bietti,~S.; Fedorov,~A.; N\"otzel,~R.; Sanguinetti,~S.
  Ehrlich-Schw\"obel effect on the growth dynamics of GaAs(111)A surfaces.
  \emph{Phys. Rev. Materials} \textbf{2017}, \emph{1}, 024602\relax
\mciteBstWouldAddEndPuncttrue
\mciteSetBstMidEndSepPunct{\mcitedefaultmidpunct}
{\mcitedefaultendpunct}{\mcitedefaultseppunct}\relax
\EndOfBibitem
\end{mcitethebibliography}


\appendix

\end{document}